\documentclass{article}

\usepackage{PRIMEarxiv}
\usepackage{url}

\usepackage[utf8]{inputenc} 
\usepackage[T1]{fontenc}   
\usepackage{hyperref}       
\usepackage{url}            
\usepackage{booktabs}       
\usepackage{amsfonts}       
\usepackage{nicefrac}       
\usepackage{microtype}      
\usepackage{lipsum}
\usepackage{fancyhdr}       
\usepackage{graphicx}       
\graphicspath{{media/}}     
\usepackage{amsmath}
\usepackage{amssymb} 
\allowdisplaybreaks 
\title{MindFlow: A Network Traffic Anomaly Detection Model Based on MindSpore
\thanks{\textit{\underline{Citation}}: 
\textbf{Authors. Title. Pages.... DOI:000000/11111.}} 
}

\author{
  Qiuyan Xiang \\
  School of Information Technology \\
  Guangxi Police College   \\
  530028,China \\
  \texttt{3310263026@qq.com} \\
   \And
  Shuang Wu \\
  College of Computer Science and AI \\
  Southwest Minzu University   \\
  610213,China \\
  \texttt{ws18774555387@outlook.com} \\
   \And
  Dongze Wu \\
  Institute of Software \\
  Chinese Academy of Sciences   \\
  100190,China \\
  \texttt{dongze@isrc.iscas.ac.cn} \\
  \And
  Yuxin Liu \\
  School of Information Technology \\
  Guangxi Police College   \\
  530028,China \\
  \texttt{3105472417@qq.com} 
  \And
  Zhenkai Qin$^{1,2,3}$ \thanks{These authors contributed equally to this work.} \\
  $^{1}$School of Computing and Information \\
  $^{2}$Network Security Research Center \\
  $^{3}$Big Data and Policing Technology Laboratory\\
  Guangxi Police College\\
  Nanning, Guangxi, China \\
  \texttt{qinzhenkai@gxjcxy.edu.cn} \\
}

\begin{document}
\maketitle

\begin{abstract}
With the wide application of IoT and industrial IoT technologies, the network structure is becoming more and more complex, and the traffic scale is growing rapidly, which makes the traditional security protection mechanism face serious challenges in dealing with high-frequency, diversified, and stealthy cyber-attacks. To address this problem, this study proposes MindFlow, a multi-dimensional dynamic traffic prediction and anomaly detection system combining convolutional neural network (CNN) and bi-directional long and short-term memory network (BiLSTM) architectures based on the MindSpore framework, and conducts systematic experiments on the NF-BoT-IoT dataset. The experimental results show that the proposed model achieves 99\% in key metrics such as accuracy, precision, recall and F1 score, effectively verifying its accuracy and robustness in network intrusion detection.
\end{abstract}


\section{Introduction}
With the accelerated development of the Internet of Things (IoT) and Industrial Internet of Things (IIoT) technologies, the network structure is becoming increasingly complex. Survey data shows that the number of global Internet users has reached 5.5 billion in 2024 \cite{1} , the number of cyber-attacks has increased by 28\% year-on-year \cite{2} , and the scale of network traffic continues to climb, a change that poses a serious challenge to traditional security protection mechanisms. Timely detection of these anomalies is essential to ensure quality of service, avoid financial losses and maintain strong security standards \cite{3} . Network traffic data usually consists of logs that summarise the communication between network-connected devices \cite{4} , which contain a large amount of sensitive communication content and access patterns that, once maliciously accessed, can lead to information leakage or privilege abuse issues. In recent years, thanks to the continuous evolution of machine learning and deep learning technologies, data-driven network traffic anomaly detection methods have gradually become the focus of research. Among these approaches, deep learning models—such as Convolutional Neural Networks (CNN) \cite{5}, Recurrent Neural Networks (RNNs) \cite{6}, and their hybrid architectures—have markedly enhanced the accuracy and response efficiency of detection systems, owing to their superior capabilities in feature extraction and temporal sequence modeling.

This study proposes MindFlow: a multidimensional dynamic flow prediction and anomaly detection system based on the MindSpore framework. The system makes full use of MindSpore's hardware acceleration, distributed computing and dynamic graph features to improve traffic prediction accuracy and anomaly detection real-time through innovative multi-dimensional feature interaction modelling and time-dependent modelling, thus providing a more efficient and accurate solution for network traffic monitoring in IoT and industrial IoT. In the experiments, the MindFlow system combines convolutional neural network (CNN) and bi-directional long and short-term memory network (BiLSTM) to construct a network traffic anomaly detection model. Firstly, the NF-BoT-IoT dataset was used for loading, feature preprocessing and time series reconstruction, and the dataset was divided into training, validation and testing sets. Then, the corresponding model structure was designed and the loss function and optimiser were configured, and multiple rounds of training were conducted. By dynamically adjusting the classification threshold, the system optimised the F1 score on the validation set performance. After the completion of training, the model parameters with optimal performance were loaded and evaluated on the test set, and the final output of high-precision prediction results was achieved. The experimental results show that the proposed model reaches 99\% in the key indexes of accuracy, precision, recall and F1 score, which fully verifies the effectiveness and robustness of the method in network intrusion detection task.

\section{Related Work}
With the development of the Internet, the massification of devices leads to the explosive growth of Internet traffic, which poses a major challenge to the management of network resources and the guarantee of network security \cite{7}. In this context, network traffic anomaly detection, as an important means of identifying potential threats and abnormal behaviours in the network, has gradually become one of the core technologies for safeguarding the security and stable operation of computer networks. This technology is mainly used to discover abnormal communications or attacks in a timely manner by analysing the statistical characteristics and behavioural patterns in network traffic data. In recent years, thanks to the wide application of machine learning technology, network traffic anomaly detection gradually realises the transformation from rule-driven to data-driven, especially the intelligent methods represented by supervised learning and unsupervised learning, which greatly improve the accuracy and efficiency of detection \cite{8}. In response to the complex and changing network security threats, researchers have continued to explore deeply in this field and proposed a variety of effective detection algorithms and model architectures, thus continuously promoting the network traffic anomaly detection technology in the direction of more efficient and smarter development.

During the early development of network traffic anomaly detection, research has mostly focused on statistical analysis and feature extraction methods, e.g., Lv et al \cite{9} introduced the Wavelet Generalised Likelihood Ratio (WGLR) algorithm and Error Performance Detection (EPD) algorithm, which combine the wavelet transform and generalised likelihood ratio methods in order to capture fault points in real time. huang et al \cite{4} proposed a Growth Hierarchy Based Self-Organising Mapping (GHSOM) anomaly detection method to understand anomalous network traffic behaviour and provide effective classification rules.Novakov et al \cite{10} investigated the application of PCA and wavelet algorithms in network traffic anomaly detection and contributed significantly to the technological advancement in the field.Ding et al \cite{11} demonstrated the use of Principal Component Analysis (PCA) in the detection of network traffic anomalies by analysing the Traffic Matrix (TM) features. Component Analysis (PCA) in network traffic anomaly detection.Bhuyan et al \cite{12} proposed a multi-step outlier-based anomaly detection method for network-wide traffic.

In recent years, researchers have been exploring network anomaly detection strategies centred on machine learning techniques, especially the application of deep neural networks, such as CNNs and RNNs, on model architectures, which significantly improve the detection accuracy and operational efficiency of the system. For example, Wei et al \cite{13} introduced a deep learning-based hierarchical spatio-temporal feature learning approach, specifically combining convolutional neural networks (CNN) and recurrent neural networks (RNN) for cyber anomaly detection.Hao et al \cite{singh2021ml} proposed a hybrid statistical-machine learning model for real-time anomaly detection of industrial cyber-physical systems, combining the dynamic thresholding model based on SARIMA with long and short-term memory (LSTM). model combined with a Long Short-Term Memory (LSTM) model.Singh et al. provided an overview of various machine learning techniques for network traffic anomaly detection, discussing the advantages and disadvantages of different models and their accuracy levels. In addition, Liu et al \cite{15} developed a real-time anomaly detection system based on convolutional neural networks for online packet extraction and analysis.





\section{Model Description}

\begin{figure}[htbp]
  \centering
  \includegraphics[width=1.0\linewidth]{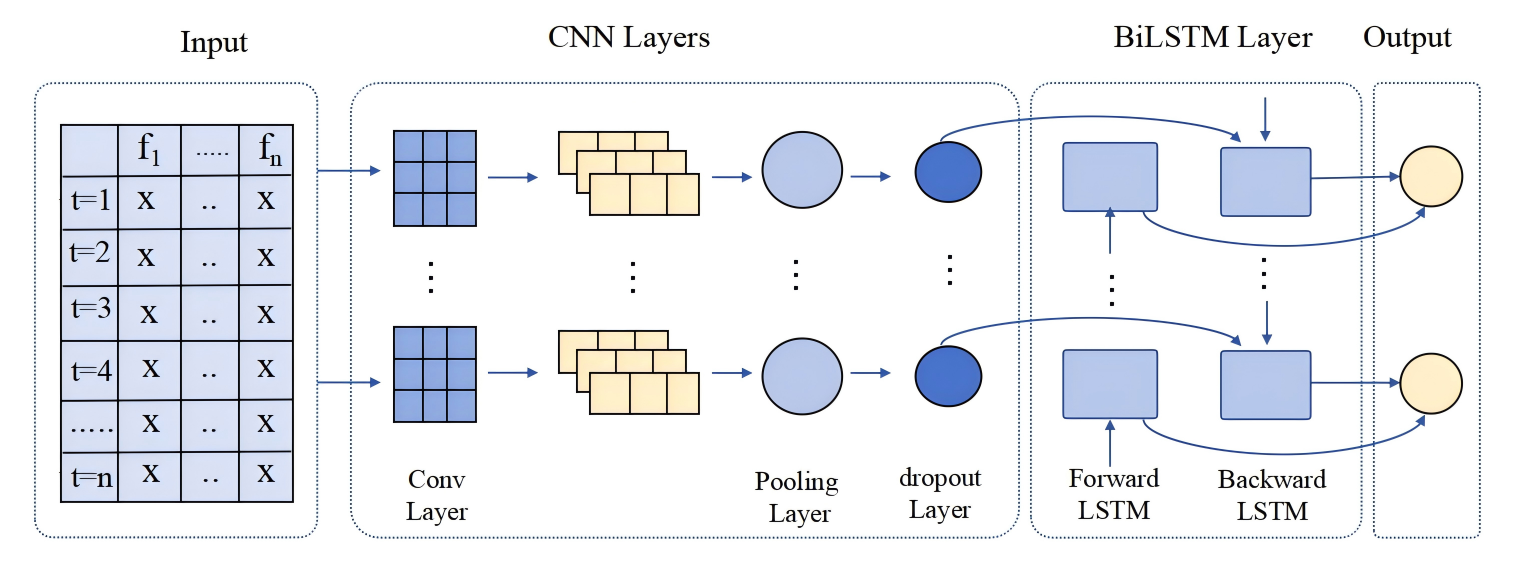}
  \caption{The overall architecture of the proposed CNN-BiLSTM network traffic anomaly detection model based on the MindSpore framework. The model integrates convolutional layers for local feature extraction and a bidirectional LSTM for temporal dependency modeling, enabling effective identification of complex spatiotemporal patterns in network traffic data.}
  \label{fig:model architecture}
\end{figure}

This model adopts the CNN-BiLSTM architecture, which combines the advantages of convolutional neural networks in local spatial feature extraction with the capability of bidirectional LSTM in temporal modelling. The specific model architecture is illustrated in Fig\ref{fig:model architecture}.
.In particular, the CNN layer mainly focuses on capturing spatial features in traffic data, while the BiLSTM layer further explores its temporal dependencies \cite{16} to achieve collaborative modelling of spatial and temporal information. This structure is particularly suitable for analysing network traffic data under attack, and can effectively identify complex spatio-temporal feature patterns, thus improving the accuracy and robustness of abnormal behaviour detection. The model first takes the network traffic feature sequences in a fixed time window as input, extracts the local patterns through multi-layer 1D convolution and pooling, and prevents overfitting through a Dropout layer; then the convolution output sequences are fed into a BiLSTM layer, which captures the context-dependent information in the forward and backward directions, and then outputs the prediction results through the fully-connected layer and the activation function. The structure can take into account the spatial locality and temporal correlation of traffic data, and shows good performance and robustness in the anomaly detection task.
\subsection{CNN}
Convolutional neural networks (CNNs) were originally introduced by Yann LeCun et al. in 1989 \cite{17}, aiming to process datasets with a grid structure, such as images and time-series data, through a weight sharing mechanism. In network traffic anomaly detection tasks, network flow data often contains rich local behavioural features, such as packet frequency and protocol fluctuations over a short period of time. By extracting traffic time slices frame by frame, CNN is able to identify potential anomaly patterns and improve the expressive power of pre-feature coding. In this experiment, a one-dimensional convolutional structure (Conv1D) is used to extract features from the reconstructed time series, and with the help of multi-layer convolution and pooling operations, the sensitivity of the model to local changes is enhanced to provide structured input for subsequent time-series modelling.

Given an input feature sequence matrix $\mathbf{X} \in \mathbb{R}^{T \times n}$, where $T$ denotes the number of time steps and $n$ represents the feature dimension at each time step, a one-dimensional convolution operation is applied by the CNN as follows:

\begin{equation}
y_i = \sum_{j=0}^{k-1} \mathbf{x}_{i+j} \cdot \mathbf{w}_j + \mathbf{b}
\end{equation}

where $\mathbf{x}_{i+j}$ denotes the feature at position $i+j$ in the input sequence; $\mathbf{w}_j$ is the convolution kernel parameter; $k$ is the kernel size;$\mathbf{b}$ is the bias term; and $y_i$ is the output feature after convolution.

\subsection{BiLSTM}
\begin{figure}[htbp]
  \centering
  \includegraphics[width=1.0\linewidth]{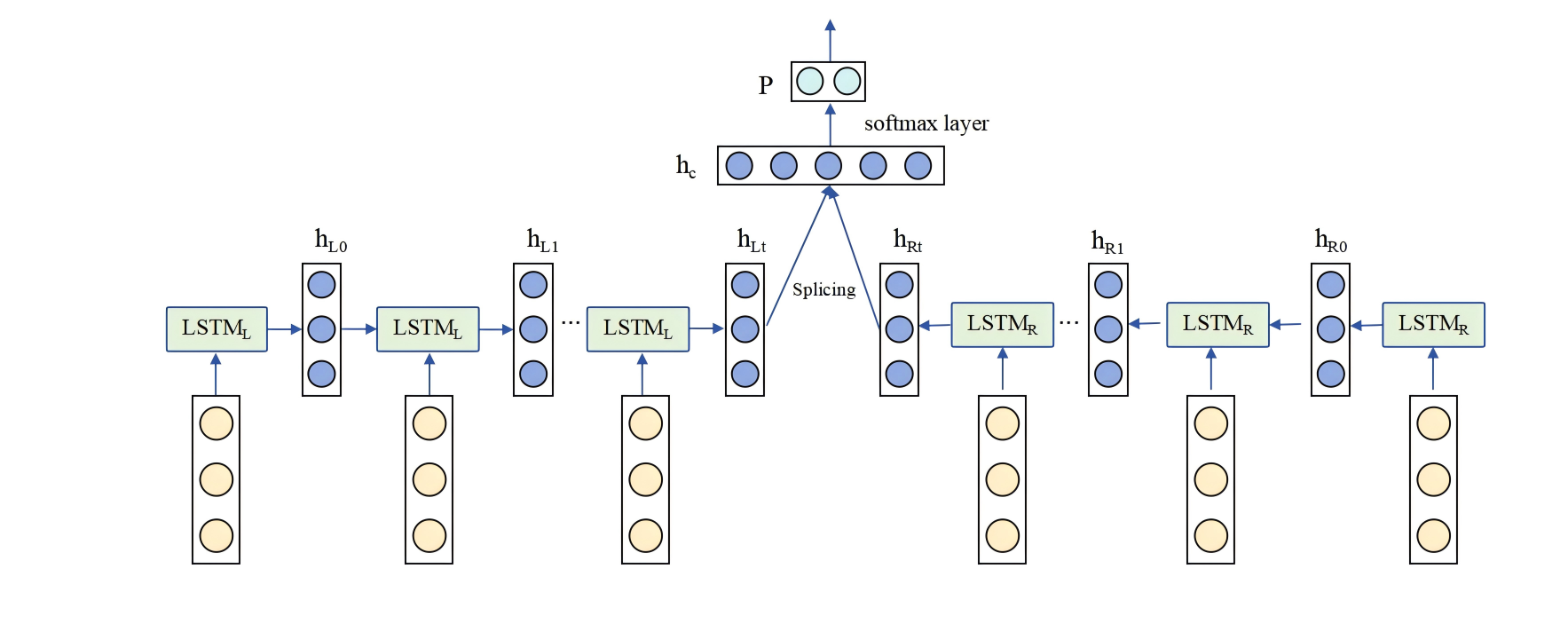} 
  \caption{Context-aware representation learning using a bidirectional LSTM. The left-to-right and right-to-left LSTM layers encode the input sequence into $\mathbf{h}{Lt}$ and $\mathbf{h}{Rt}$, which are concatenated to form the context vector $\mathbf{h}_c$. The final prediction $\mathbf{P}$ is obtained via a softmax layer.}
  \label{fig:BiLSTM model architecture}
\end{figure}

Bidirectional Long Short-Term Memory\cite{18} was originally proposed by Alex Graves and Jürgen Schmidhuber in 2005, with the aim of considering both forward and backward contextual information in sequence modelling to improve the model's ability to model global sequence dependencies.LSTM has excellent long-term dependency modelling capabilities by introducing forgetting gates, LSTM has excellent long-term dependency modelling ability by introducing the mechanisms of forgetting gate, input gate and output gate, effectively retaining important historical information and suppressing irrelevant interference.The specific model architecture is illustrated in Fig\ref{fig:BiLSTM model architecture}. In this experiment, the BiLSTM network is used to jointly model the forward and backward time-series characteristics of network traffic, which improves the recognition ability and detection accuracy of abnormal behaviour.

First, the input sequence is processed in chronological order to compute the forward hidden state at each time step, which is used to extract forward temporal features:

\begin{equation}
\overrightarrow{h}_t = \text{LSTM}_L(x_t, \overrightarrow{h}_{t-1}), \quad t = 0, 1, \ldots, T
\end{equation}

Then, the input sequence is processed in reverse chronological order to compute the backward hidden state at each time step, which is used to extract backward temporal features:

\begin{equation}
\overleftarrow{h}_t = \text{LSTM}_R(x_t, \overleftarrow{h}_{t+1}), \quad t = T, T-1, \ldots, 0
\end{equation}

Finally, the forward and backward hidden states at each time step are concatenated to form a temporal representation that integrates past and future information:

\begin{equation}
h_c = \left[ \overrightarrow{h}_T ; \overleftarrow{h}_T \right]
\end{equation}

\noindent
where:
\begin{itemize}
    \item $t$ denotes the index of the current time step, where $0 \leq t \leq T$;
    \item $x_t$ represents the word embedding of the input sequence at time step $t$;
    \item $\overrightarrow{h_t}$ denotes the hidden state of the forward LSTM at time step $t$;
    \item $\overleftarrow{h_t}$ denotes the hidden state of the backward LSTM at time step $t$;
\end{itemize}

\section{Experimental Results and Analysis}
\subsection{Dataset}
The NF-BoT-IoT dataset\cite{19} was meticulously developed by the University of Queensland based on the NetFlow format and represents a specialized dataset for Internet of Things (IoT) network security research. Its raw data originates from the BoT-IoT dataset, constructed by the Cyber Range Lab at the University of New South Wales Canberra (UNSW Canberra), Australia. The NF-BoT-IoT dataset comprises approximately 600,000 network flow records, among which only 2.31\% represent benign traffic, while a substantial 97.69\% correspond to malicious traffic, including various types of attacks such as reconnaissance, DDoS, DoS, and data theft. Owing to its comprehensive coverage of attack types and clearly labeled data, the dataset has been widely adopted as a standard benchmark in intrusion detection systems (IDS) and cybersecurity research. It provides a reliable and reproducible experimental platform, thereby significantly advancing the development of related studies.
\subsection{Experimental Environment}
The experiments were conducted on a local computing environment running Ubuntu 22.04. The hardware configuration includes an 8-core CPU and 32 GB of RAM. The software environment is based on Python 3.10, with core dependencies on the MindSpore deep learning framework, as well as common scientific computing libraries such as NumPy, pandas, and scikit-learn. The development platform is pre-installed with the ModelScope Library, supporting rapid development and seamless environment integration.
\subsection{Evaluation Metrics}
In this paper, Accuracy, Precision, Recall and F1 score are used as the main evaluation indexes of model performance. Among them, Accuracy is used to measure the proportion of samples on which the model's prediction results are consistent with the true labels, which is a basic indicator of the model's overall prediction ability. The higher the accuracy rate, the closer the model's classification results on all samples are to the real situation. Its specific calculation formula is as follows:

\begin{equation}
\text{Accuracy} = \frac{TP + TN}{TP + TN + FP + FN}
\end{equation}
\vspace{0.5em}
\begin{equation}
\text{Precision} = \frac{TP}{TP + FP}
\end{equation}
\vspace{0.5em}
\begin{equation}
\text{Recall} = \frac{TP}{TP + FN}
\end{equation}
\vspace{0.5em}
\begin{equation}
F1 = 2 \times \frac{\text{Precision} \times \text{Recall}}{\text{Precision} + \text{Recall}}
\end{equation}
\vspace{0.5em}

Among them, $TP$ denotes the number of positive samples correctly identified by the model, $TN$ represents the number of negative samples correctly identified, $FP$ refers to the number of negative samples incorrectly classified as positive, and $FN$ indicates the number of positive samples misclassified as negative. These metrics are used to evaluate the model's capability to distinguish between positive and negative classes.


\subsection{Experimental Results and Analysis}

To comprehensively assess the effectiveness of the proposed CNN-BiLSTM-based anomaly detection model, MindFlow, we conducted extensive experiments using the NF-BoT-IoT dataset. 

\begin{figure}[htbp]
  \centering
  \includegraphics[width=0.75\linewidth]{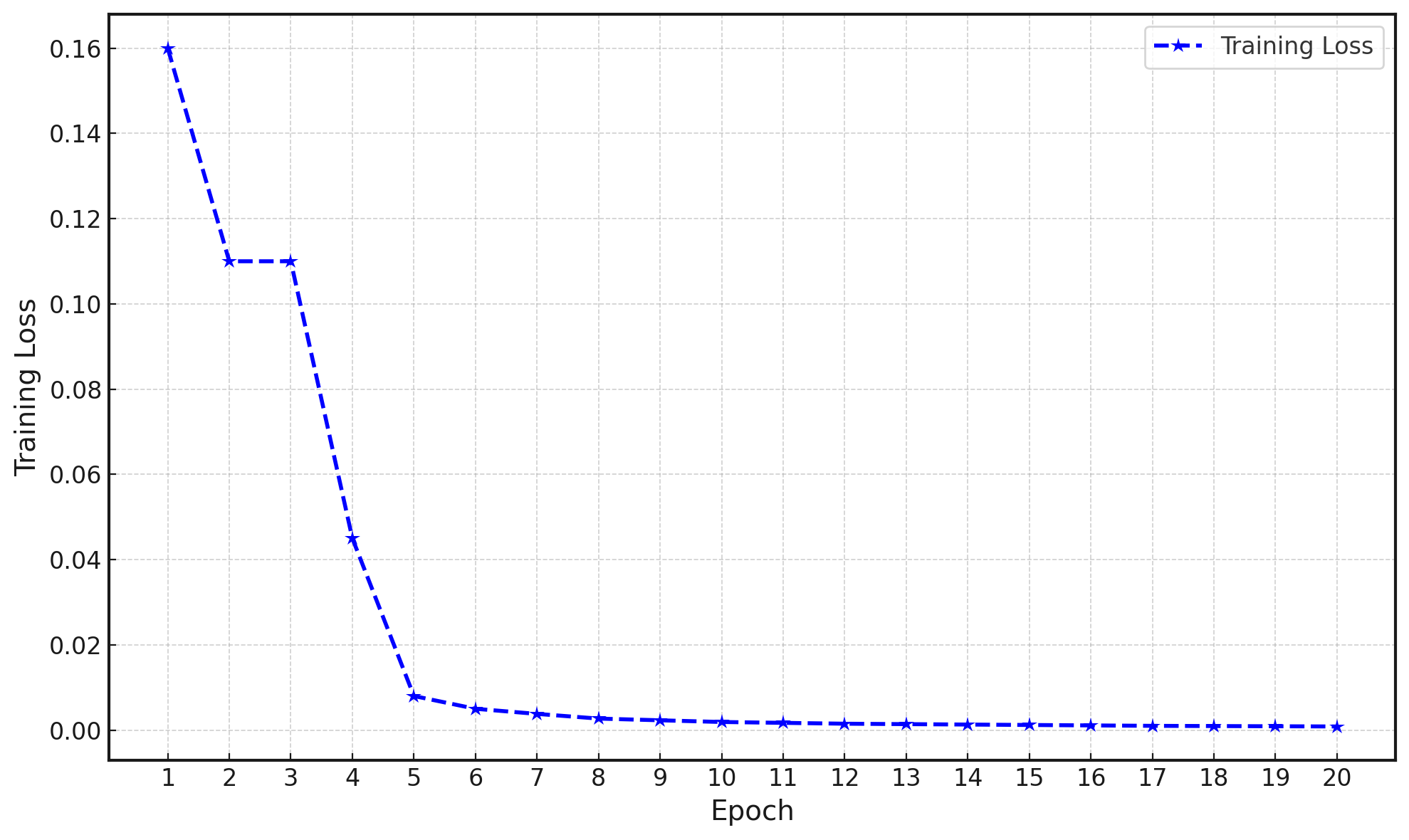}
  \caption{Training loss curve of the CNN-BiLSTM model over 20 epochs. The loss decreased from 0.1650 to 0.0002, indicating rapid convergence and stable optimization.}
  \label{fig:training_loss}
\end{figure}

The model was trained over 20 epochs. As shown in Fig.~\ref{fig:training_loss}, the training loss rapidly decreased from an initial value of 0.1650 to 0.0002, indicating efficient convergence and stable training dynamics. The sharp loss reduction in the early epochs suggests that the model quickly captures key discriminatory features from the input sequences. The subsequent plateau phase indicates convergence, with no signs of overfitting, underscoring the model's learning efficiency and generalization capability.

This rapid convergence is largely attributable to the complementary strengths of the CNN and BiLSTM components. The CNN layers are adept at extracting spatially localized patterns in network flow data—such as packet size fluctuations and short-term protocol anomalies—while the BiLSTM layers model bidirectional temporal dependencies, which are essential for recognizing complex, multi-stage attack behaviors over time.

Evaluation on the test subset of NF-BoT-IoT reveals consistently high performance, with the model achieving 99\% in all key classification metrics, including accuracy, precision, recall, and F1-score. These metrics reflect the model’s excellent ability to distinguish between benign and malicious traffic, with minimal false positives and false negatives. Such reliability is crucial for real-world IoT applications, where timely and accurate anomaly detection is vital.

Moreover, the implementation of MindFlow on the MindSpore framework further enhances its practicality. MindSpore's support for dynamic computation graphs and hardware-level acceleration contributes to reduced training time and lower resource consumption, making the system viable for deployment in edge computing scenarios or resource-constrained environments.

In conclusion, the experimental findings validate MindFlow as a highly effective model for real-time network anomaly detection. Its rapid training convergence, superior detection performance, and efficient resource utilization collectively position it as a promising solution for next-generation cybersecurity infrastructures.

\section{Conclusion}
Aiming at the problem of insufficient detection accuracy of traditional security mechanisms in the complex network environments of IoT and Industrial IoT, this study proposes MindFlow, a multi-dimensional dynamic traffic prediction and anomaly detection system combining Convolutional Neural Networks (CNNs) and Bidirectional Long and Short-Term Storage Memory Networks (BiLSTMs) architectures based on the MindSpore framework and conducts systematic experiments on the NF-BoT-IoT dataset. set on which system experiments were conducted. The experimental results show that the accuracy, precision, recall and F1 score of the model on the NF-BoT-IoT dataset are all over 99\%, demonstrating excellent detection performance and strong generalisation capability. Although the model performs stably on a single dataset, it still faces challenges such as cross-domain adaptation and real-time optimisation in practical deployment, and future research can further combine techniques such as lightweight network structure and federated learning to improve the model's adaptability and deployment efficiency in multi-scenario environments.

\section*{Acknowledgments}
Thanks for the support provided by the MindSpore Community.
\bibliographystyle{unsrt}  
\bibliography{references}

\begin{thebibliography}{10}

\bibitem{1}
J.~D’Souza.
\newblock Internet statistics by demographics, users, technologies and traffic.
\newblock \url{https://www.coolest-gadgets.com/internet-statistics/}, 2025.
\newblock Accessed: 2025-04-09.

\bibitem{2}
M.~Praharaj.
\newblock Shifting attack landscapes and sectors in q1 2024 with a 28\% increase in cyber attacks globally.
\newblock \url{https://businessnewsthisweek.com/business/shifting-attack-landscapes-and-sectors-in-q1-2024-with-a-28percent-increase-in-cyber-attacks-globally/}.
\newblock 2024. Accessed: 2025-04-09.

\bibitem{3}
Mahshid Rezakhani, Tolunay Seyfi, and Fatemeh Afghah.
\newblock A transfer learning framework for anomaly detection in multivariate iot traffic data.
\newblock {\em arXiv preprint arXiv:2501.15365}, 2025.

\bibitem{4}
Shin-Ying Huang and Yen-Nun Huang.
\newblock Network traffic anomaly detection based on growing hierarchical som.
\newblock In {\em 2013 43rd Annual IEEE/IFIP International Conference on Dependable Systems and Networks (DSN)}, pages 1--2. IEEE, 2013.

\bibitem{5}
Nelly Elsayed, Zaghloul~Saad Zaghloul, Sylvia~Worlali Azumah, and Chengcheng Li.
\newblock Intrusion detection system in smart home network using bidirectional lstm and convolutional neural networks hybrid model.
\newblock In {\em 2021 IEEE international midwest symposium on circuits and systems (MWSCAS)}, pages 55--58. IEEE, 2021.

\bibitem{6}
Benjamin~J Radford, Bartley~D Richardson, and Shawn~E Davis.
\newblock Sequence aggregation rules for anomaly detection in computer network traffic.
\newblock {\em arXiv preprint arXiv:1805.03735}, 2018.

\bibitem{7}
Zhangxuan Dang, Yu~Zheng, Xinglin Lin, Chunlei Peng, Qiuyu Chen, and Xinbo Gao.
\newblock Semi-supervised learning for anomaly traffic detection via bidirectional normalizing flows.
\newblock {\em arXiv preprint arXiv:2403.10550}, 2024.

\bibitem{8}
Wasim~A Ali, KN~Manasa, Malika Bendechache, Mohammed Fadhel~Aljunaid, and P~Sandhya.
\newblock A review of current machine learning approaches for anomaly detection in network traffic.
\newblock {\em Journal of Telecommunications and the Digital Economy}, 8(4):64--95, 2020.

\bibitem{9}
Jun Lv, Xing Li, and Tong Li.
\newblock Web-based application for traffic anomaly detection algorithm.
\newblock In {\em Second International Conference on Internet and Web Applications and Services (ICIW'07)}, pages 44--44. IEEE, 2007.

\bibitem{10}
Stevan Novakov, Chung-Horng Lung, Ioannis Lambadaris, and Nabil Seddigh.
\newblock Studies in applying pca and wavelet algorithms for network traffic anomaly detection.
\newblock In {\em 2013 IEEE 14th International Conference on High Performance Switching and Routing (HPSR)}, pages 185--190. IEEE, 2013.

\bibitem{11}
Meimei Ding and Hui Tian.
\newblock Pca-based network traffic anomaly detection.
\newblock {\em Tsinghua Science and Technology}, 21(5):500--509, 2016.

\bibitem{12}
Monowar~H Bhuyan, Druba~K Bhattacharyya, and Jugal~K Kalita.
\newblock A multi-step outlier-based anomaly detection approach to network-wide traffic.
\newblock {\em Information Sciences}, 348:243--271, 2016.

\bibitem{13}
Guanglu Wei and Zhonghua Wang.
\newblock Adoption and realization of deep learning in network traffic anomaly detection device design.
\newblock {\em Soft Computing}, 25(2):1147--1158, 2021.

\bibitem{singh2021ml}
Richa Singh, Nidhi Srivastava, and Ashwani Kumar.
\newblock Machine learning techniques for anomaly detection in network traffic.
\newblock In {\em 2021 sixth international conference on image information processing (ICIIP)}, volume~6, pages 261--266. IEEE, 2021.

\bibitem{15}
Haitao Liu and Haifeng Wang.
\newblock Real-time anomaly detection of network traffic based on cnn.
\newblock {\em Symmetry}, 15(6):1205, 2023.

\bibitem{16}
Prathamesh Chandekar, Mansi Mehta, and Swet Chandan.
\newblock Enhanced anomaly detection in iomt networks using ensemble ai models on the ciciomt2024 dataset.
\newblock {\em arXiv preprint arXiv:2502.11854}, 2025.

\bibitem{17}
Yann LeCun, L{\'e}on Bottou, Yoshua Bengio, and Patrick Haffner.
\newblock Gradient-based learning applied to document recognition.
\newblock {\em Proceedings of the IEEE}, 86(11):2278--2324, 1998.

\bibitem{18}
Alex Graves and J{\"u}rgen Schmidhuber.
\newblock Framewise phoneme classification with bidirectional lstm and other neural network architectures.
\newblock {\em Neural networks}, 18(5-6):602--610, 2005.

\bibitem{19}
Nf-bot-iot - uq espace.
\newblock \url{https://espace.library.uq.edu.au/view/UQ:62e6d80}, 2025.
\newblock Accessed: 2025-04-09.

\end{thebibliography}

\end{document}